# Inversion Symmetry Breaking in Noncollinear Magnetic Phase of a Triangular Lattice Antiferromagnet $CuFeO_2$

T. Kimura,[1] J. C. Lashley,[1] and A. P. Ramirez[2]

[1]*Los Alamos National Laboratory, Los Alamos, NM 87545, USA*

[2]*Bell Laboratories, Lucent Technologies, 600 Mountain Avenue, Murray Hill, NJ 07974, USA*

Magnetoelectric and magnetoelastic phenomena correlated with a phase transition into noncollinear magnetic phase have been investigated for single crystals of $CuFeO_2$ with a frustrated triangular lattice. $CuFeO_2$ exhibits several long-wavelength magnetic structures related to the spin frustration, and it is found that finite electric polarization, namely inversion symmetry breaking, occurs with noncollinear but not at collinear magnetic phases. This result demonstrates that the noncollinear spin structure is a key role to induce electric polarization, and suggests that frustrated magnets which often favor noncollinear configurations can be plausible candidates for magnetoelectrics with strong magnetoelectric interaction.

In recent years, geometrically frustrated magnetic systems have received a great deal of attention because their ordered or unordered ground states are often very exotic[1,2,3]. The triangular lattice antiferromagnet (TLA); in which magnetic ions reside at a triangular net, is the most obvious example of a geometrically frustrated magnetic system. A typical ordered structure in TLA is a noncollinear 3-sublattice 120° spin configurations where the frustration of the three nearest-neighbor spins on a triangular plaquette is resolved by a 120° rotation of neighboring spins. The $ABO_2$ family with the *delafossite* structure ($A$ = monovalent ion, $B$ = trivalent ion) has been recently investigated as one of the typical materials for the TLA. Indeed, some of them such as $LiCrO_2$ and $CuCrO_2$ demonstrate the 120° spin structure with a weak easy-axis anisotropy[2]. Among delafossite compounds, $CuFeO_2$ which is a naturally occurring mineral was historically the first known compound exhibiting the crystal structure[4]. The fundamental crystal structure of $CuFeO_2$ (right upper inset of Fig. 1) belongs to the space group $R\bar{3}m$ and consists of a triangular lattice of magnetic $Fe^{3+}$ ions ($S$=5/2, $L$=0; orbital singlet) separated by nonmagnetic ionic layers of $O^{2-}$, $Cu^{+}$, and $O^{2-}$, stacked along the $c$ axis in the hexagonal description[5]. Among other magnetic delafossite $ABO_2$ compounds, $CuFeO_2$ has unique magnetic properties in which a collinear commensurate 4-sublattice (↑↑↓↓) magnetic structure [collinear-CM(1/4) phase] with magnetic moments along the $c$ axis [characteristic modulation wave vector: ($h$ $h$ 1/2) where $h$ = 1/4] is realized in each layer at the zero-field ground state[6] (left lower inset of Fig. 1). With increasing temperature $T$, the system shows a sinusoidally amplitude-modulated incommensurate structure $T_{N2}$~11 K where the magnetic moments are collinearly coupled and the modulation wave number $h$ is $T$-dependent (~1/4 > $h$ > ~1/5) [collinear-ICM phase], and then becomes paramagnetic at $T_{N1}$~14 K[7,8].

One of the most intriguing properties of $CuFeO_2$ is the evolution of its magnetic structures, i.e. multistep metamagnetic transitions, when a magnetic field *B* is applied along the *c* axis[9,10]. Figure 1 displays the magnetic phase diagram of $CuFeO_2$ crystals we investigated (The phase boundaries were determined from anomalies in magnetization, dielectric constant, electric polarization, and striction.), which bears a close resemblance to that presented in refs. [11,12]. The application of *B* between ~13 T and ~20 T induces a collinear-CM(1/5) phase in which collinear moments along the *c* axis in each layer exhibit the (↑↑↑↓↓) configuration where *h* = 1/5, as illustrated in the lower right inset of Fig. 1. Between the collinear-CM(1/4) and collinear-CM(1/5) phases (~7 T < *B* < ~13 T) there exists the first *B*-induced state (noncollinear-ICM phase in Fig. 1), which is the focus of this paper. A proposed model for the magnetic structure of the first *B*-induced state based on a neutron diffraction measurement is the twisted helical structure where the magnetic moments rotate in a twisted helical manner and noncollinearly align along the <110> direction and couple antiferromagnetically between adjacent *c* layers. In addition, the modulation wave number *h* for the helical structure is incommensurate and *B*-dependent (~1/4 > *h* > ~1/5). Although the multistep metamagnetic feature has been interpreted in terms of the 2 dimensional Ising model[8], the 3D Heisenberg model can explain the appearance of the *B*-induced noncollinear phase[9].

In this paper, we discuss another intriguing aspect, namely *multiferroic* nature, of geometrical spin frustration in a TLA. Recently, there has been a revival of interest in research of multiferroics showing the coexistence and/or interplay of magnetism and ferroelectricity so-called magnetoelectric (ME) effects[13,14]. Among multiferroics studied to date, some exhibit strong couplings between magnetism and ferroelectricity as well as long-wavelength magnetic structures[15,16,17,18,19]. Synchrotron x-ray diffraction studies of one compound, $TbMnO_3$, reveal the appearance of magnetoelastically-induced lattice distortion with nonzero wave vector and its lock-in transition (or ICM-CM transition) accompanied by ferroelectric (FE) order[16]. This highlights the importance of the ICM-CM transition or the appearance of a CM phase which causes FE order in some improper ferroelectrics (e.g. $Rb_2ZnCl_4$)[20]. However, a recent theoretical study by Katsura and co-workers[21] pointed out a possible microscopic mechanism of the ME effect in noncollinear magnets based on the spin supercurrent, which suggests that noncollinear magnets with *spiral* spin structure should show finite electric polarization. Furthermore, a recent neutron diffraction measurement on $TbMnO_3$ proposed that the FE phase is accompanied by a transversely-modulated spiral magnetic structure[22]. These studies indicate that the noncollinear spin structure with spin helicity is a key to understand inversion symmetry breaking in multiferroics showing long-wavelength magnetic structures. Here, we show that inversion symmetry can be broken only at a noncollinear magnetic phase in a TLA, $CuFeO_2$.

We have grown single crystals of $CuFeO_2$ by the floating zone method, following Ref. [23]. The crystals were oriented using Laue x-ray diffraction patterns, and cut into thin plates with the widest faces parallel and perpendicular to the *c* axis in the hexagonal setting. Gold electrodes were then vacuum-deposited onto these faces for measurements of dielectric constant ε and electric polarization *P*. We measured ε at 1 MHz using an LCR meter, and obtained *P* by measurements of the pyroelectric (or ME) current with varying *T* (or *B*). Before the measurements of *P*, a proper ME cooling process was performed to obtain a single FE domain. The magnetization *M* and ac susceptibility χ' were measured with a commercial magnetometer. The magnetostriction *L* was measured using uniaxial strain gauges which were attached to the widest face of the specimens. The contribution of the gauge's magnetoresistance [$\Delta\rho(B)/\rho(0) \sim B^2$] was subtracted after the measurements.

Figure 2(a) displays the temperature profiles of ac susceptibility parallel ($\chi'_{//}$) and perpendicular ($\chi'_{\perp}$) to the $c$ axis in an ac magnetic field of 1 mT and 1 kHz. $\chi'_{//}$ shows a broad maximum at $T_{N1}$~14 K and then suddenly drops at $T_{N2}$~11 K on cooling. The anomalies in $\chi'$ at $T_{N1}$ and $T_{N2}$ are associated with the transitions from paramagnetic to collinear-ICM phase and from collinear-ICM to collinear-CM(1/4) phases, respectively. Measurements of ε revealed that the magnetic phase transitions strongly affect dielectric properties, as shown in Fig. 2(b). The temperature profiles of ε with an electric field parallel ($\varepsilon_{//}$) and perpendicular ($\varepsilon_{\perp}$) to the $c$ axis[24] show distinct anomalies at $T_{N1}$ and $T_{N2}$. Both $\varepsilon_{//}$ and $\varepsilon_{\perp}$ show a rapid increase at $T_{N1}$ toward lower temperatures, and form peak structures with a peak centered at ~12 K. With further decreasing $T$, $\varepsilon_{//}$ shows a sudden jump while $\varepsilon_{\perp}$ steeply decreases at $T_{N2}$. In addition, the dielectric anomaly around $T_{N2}$ is accompanied by a substantial thermal hysteresis, which suggests a first order phase transition at $T_{N2}$. We also investigated the $T$ dependence of $P$ parallel ($P_{//}$) and perpendicular ($P_{\perp}$) to the $c$ axis by measuring the pyroelectric current. However, no substantial pyroelectric current was detected in either directions. This shows that the ordered magnetic states at zero magnetic field [collinear-ICM and collinear-CM(1/4) phases] cannot induce the inversion symmetry breaking though the dielectric anomaly is evident at the transitions.

To examine how the evolution of magnetic ordered states affects electric properties, we show in Fig. 3 the pyroelectric coefficient perpendicular to the $c$ axis ($p_{\perp}$) at selected magnetic fields along the $c$ axis as a function of $T$. No substantial pyroelectric coefficient has been detected at $B$ < ~5 T where the collinear-ICM phase is not realized at any temperature. However, when $B$ is applied at 6 T, two anomalies of $p_{\perp}$ occur at 10.5 and 10.9 K. The opposite sign of these anomalies indicates that the polar nature appears between the two temperatures. With increasing $B$, the both low-$T$ and high-$T$ anomalies are shifted toward lower T. Above 8 T, only the low-$T$ anomaly vanishes, which means the polar phase persists down to the lowest temperature. With further increasing $B$, the high-$T$ anomaly rapidly shifts toward lower $T$, and disappears at ~14 T where substantial $p_{\perp}$ is not observed in any $T$ range as at $B$ < ~5 T. Comparing the data in Fig. 3 and magnetic phase diagram shown in Fig. 1, the $B$-induced polar phase perfectly coincides with the noncollinear-ICM phase.

To further verify the relation between the magnetic and electric properties, we display the $B$-dependence of $M$ and $P_{\perp}$ at selected temperatures in Figs. 4(a) and 4(b), respectively. As seen in Fig. 4(a), two magnetization steps were observed at $B \leq 14$ T and $T \leq 10$ K. The first and the second steps correspond to the transitions into the noncollinear-ICM and the collinear-CM(1/5) phases, respectively. The metamagnetic features vanish at 11 K where the collinear-ICM state is stabilized. Comparison of the $M$-$B$ curves with the $P$-$B$ curves at the respective $T$s reveals the strong interplay of $M$ and $P$. As mentioned above, inversion symmetry is preserved at the collinear-CM(1/4) phase. The data in lower $B$ region of Fig. 4(b) clearly show that $P$ is not induced by $B$ at the collinear-CM(1/4) phase. However, $P$ exhibits a sudden increase at the transition field into the noncollinear-ICM phase, and becomes finite[25]. We also confirmed the sign reversal of $P_{\perp}$ by reversing poling electric fields. The magnitude of $P_{\perp}$ (~$10^2$ μC/m$^2$) at the noncollinear-ICM phase is comparable to those observed in known multiferroics with long-wavelength magnetic structures[15-18]. With further increasing $B$, $P$ vanishes again at the transition field into the collinear-CM(1/5) phase. It should be emphasized that $P_{\perp}$ becomes finite only at the noncollinear-ICM phase but not at the collinear-CM and collinear-ICM phases. In addition, it is also worth mentioning that inversion symmetry is broken at an *incommensurate* phase not at commensurate ones in $CuFeO_2$, unlike in conventional improper ferroelectrics where FE order emerges at a phase transition from an incommensurate to a commensurate phase[20]. This result clearly demonstrates that the appearance of a noncollinear magnetic structure plays a key role in breaking the inversion symmetry of multiferroics with long-wavelength magnetic structures.

Let us also mention the lattice distortion accompanied by the magnetic and magnetoelectric phase transitions. Figures 4(c) and 4(d) show the isothermal magnetostriction parallel [$\Delta L_{//}(B)/L_{//}(0)$] and perpendicular [$\Delta L_{\perp}(B)/L_{\perp}(0)$] to the *c* axis at selected temperatures. The magnetostriction is highly anisotropic between $\Delta L_{//}(B)/L_{//}(0)$ and $\Delta L_{\perp}(B)/L_{\perp}(0)$. Comparing the magnetostriction with *M* and *P* for the respective temperatures, one may notice a close interrelation among the magnetic, magnetoelectric, and magnetoelastic properties. At 11 K where metamagnetic transition does not occur, no remarkable magnetostriction has been observed. However, below 10 K, switching-like large magnetostriction takes place at onset fields of magnetic and magnetoelectric transitions. At 10 K where the three phases [collinear-CM(1/4), noncollinear-ICM, and collinear-CM(1/5) phases] can be realized below 9 T, two steps have been observed in the magnetostriction at their phase boundaries. In the both steps, $\Delta L_{//}(B)/L_{//}(0)$ abruptly increases while $\Delta L_{\perp}(B)/L_{\perp}(0)$ decreases toward higher-*B* induced phases. In data below 9 K, only one step can be seen at the transition from collinear-CM(1/4) to noncollinear-ICM phases since *B* of 9 T is not enough to induce collinear-CM(1/5) phase. Thus, as the system undergoes phase transitions into higher-*B* phases, the *c* axis elongates while the *ab* plane shrinks.

The observed magnetostriction can be originated from the change of nearest-neighboring Fe-O-Fe bond angle ($\phi$) as well as Fe-O length in the delafossite structure (See the right inset of Fig. 1). In the fundamental crystal structure of $CuFeO_2$, the $\phi$ is ~96.7° which is rather close to 90°.[5] The Goodenough-Kanamori rules[26 27 28] suggest that the 180° superexchange $d^5$-$d^5$ interaction has strong antiferromagnetic coupling while that of the 90° interaction is uncertain or weakly antiferromagnetic because of canceling ferromagnetic and antiferromagnetic effects. The striction behavior, i.e. the elongation of the *c* axis and the reduction of the *ab* plane, may be caused by the decrease of average $\phi$. As the system undergoes metamagnetic transitions with increasing *B*, the ratio of ferromagnetically-coupled nearest-neighboring Fe sites (*f*) increases [e.g. $f = 1/3$ in collinear-CM(1/4) phase, $f = 7/15$ in collinear-CM(1/5) phase]. It is possible to consider that the increase of *f* gives rise to the decrease of average $\phi$, and then causes the elongation of the *c* axis and the reduction of the *ab* plane. The lattice distortion may somewhat relax the frustration in the TLA. However, detailed investigations of the crystallographic structures by neutron and/or synchrotron x-ray diffraction studies are needed to test this suggestion. It is worth mentioning that a strong spin-lattice coupling exists in multiferroics containing geometrical spin frustration.

In summary, we investigated the magnetic, magnetoelectric, and magnetoelastic properties of a triangular lattice antiferromagnet $CuFeO_2$ showing the magnetic-field-induced collinear-noncollinear magnetic phase transitions. The present study demonstrates that geometrically frustrated magnetic systems which often favor noncollinear magnetic structures are good candidates for multiferroics with strong magnetoelectric interaction.

We gratefully acknowledge discussions with R. Kajimoto, F. Ye, Y. Ren, and G. Lawes, and thank J. L. Sarrao and K. J. McCllelan for help with experiments. This work was supported by the U.S. DOE.

**Fig. 1. (Color online) Temperature (*T*) versus magnetic field (*B*) phase diagram of $CuFeO_2$ with *B* applied along the *c* axis. Open and filled symbols represent the data points in the cooling (or *B*-decreasing) and warming (or *B*-increasing) runs, respectively. Blue, red, black, and green data points were obtained by measurements of magnetization, dielectric constant, electric polarization, and striction, respectively. Upper inset: Crystal structure of $CuFeO_2$. Lower insets: Schematic illustrations of magnetic structures on $Fe^{3+}$ sites at (left) the collinear-CM(1/4) and (right) the collinear-CM(1/5) states. White and black circles correspond to up and down spin states, respectively. Inversion symmetry is broken at the noncollinear-ICM phase (gray area).**

**Fig. 2. Temperature profiles of (a) magnetic susceptibility parallel ($\chi_{//}$) and perpendicular ($\chi_{\perp}$) to the *c* axis and (b) dielectric constant for electric fields applied parallel ($\varepsilon_{//}$) and perpendicular ($\varepsilon_{\perp}$) to the *c* axis in $CuFeO_2$.**

**Fig. 3. (Color online) Pyroelectric coefficient as a function of temperature at selected magnetic fields for $CuFeO_2$. Magnetic fields were applied along the *c* axis, while pyroelectric current was measured in the direction perpendicular to the *c* axis.**

**Fig. 4. (Color online) Magnetization (a), electric polarization perpendicular to the *c* axis (b), and magnetostriction parallel (c) and perpendicular (d) to the *c* axis of $CuFeO_2$ as a function of magnetic field at selected temperatures. Magnetic field was applied along the *c* axis.**

[1] A. P. Ramirez, Annu. Rev. Mater. Sci. **24**, 453 (1994).

[2] M. F. Collins and O. A. Petrenko, Can. J. Phys. **75**, 605 (1997).

[3] H. Kawamura, J. Phys.: Condens. Matter **10**, 4707 (1998).

[4] A. Pabst, Amer. Mineral. **31**, 539 (1946).

[5] C. T. Prewitt, R. D. Shannon, and D. B. Rogers, Inorg. Chem. **10**, 791 (1971).

[6] S. Mitsuda, H. Yoshizawa, N. Yamaguchi, and M. Mekata, J. Phys. Soc. Jpn. **60**, 1885 (1991).

[7] S. Mitsuda, N. Kasahara, T. Uno, and M. Mase, J. Phys. Soc. Jpn. **67**, 4026 (1998).

[8] N. Terada, T. Kawasaki, S. Mitsuda, H. Kimura, and Y. Noda, J. Phys. Soc. Jpn. **74**, 1561 (2005).

[9] Y. Ajiro, T. Asano, T. Takagi, M. Mekata, H. A. Katori, and T. Goto, Physica B **201**, 71 (1994).

[10] O. A. Petrenko, M. R. Lees, G. Balakrishnan, S. de Brion, and G. Chouteau, J. Phys.: Condens. Matter **17**, 2741 (2005).

[11] S. Mitsuda, M. Mase, T. Ueno, H. Kitazawa, and H. A. Katori, J. Phys. Soc. Jpn. **69**, 33 (2000).

[12] S. Mitsuda, M. Mase, K. Prokes, H. Kitazawa, and H. A. Katori, J. Phys. Soc. Jpn. **69**, 3513 (2000).

[13] N. A. Hill, J. Phys. Chem. B **104**, 6694 (2000).

[14] M. Fiebig, J. Phys. D: Appl. Phys. **38**, R123 (2005).

[15] T. Kato, K. Machida, T. Ishii, and K. Iio, Phys. Rev. B **50**, 13039 (1994).

[16] A. Inomata and K. Kohn, J. Phys.: Condens. Matter **8**, 2673 (1996).

[17] T. Kimura, T. Goto, H. Shintani, K. Ishizaka, T. Arima, and Y. Tokura, Nature **426**, 55 (2004).

[18] T. Kimura, G. Lawes, and A. P. Ramirez, Phys. Rev. Lett. **94**, 137201 (2005).

[19] G. Lawes, A. B. Harris, T. Kimura, N. Rogado, R. J. Cava, A. Aharony, O. Entin-Wohlman, T. Yildirim, M. Kenzelmann, C. Broholm, and A. P. Ramirez, Phys. Rev. Lett. **95**, 087205 (2005).

[20] R. Blinc and A. P. Levanyuk, *Incommensurate Phases in Dielectrics 1. Fundamentals* (North-Holland, Amsterdam, 1986).

[21] H. Katsura, N. Nagaosa, and A. V. Balatsky, Phys. Rev. Lett. **95**, 057205 (2005).

[22] M. Kenzelmann, A. B. Harris, S. Jonas, C. Broholm, J. Schefer, S. B. Kim, C. L. Zhang, S.-W. Cheong, O. P. Vajk, and J. W. Lynn, Phys. Rev. Lett. **95**, 087206 (2005).

[23] T. R. Zhao, M. Hasegawa, and H. Takei, J. Cryst. Growth **166**, 408 (1996).

[24] We measured the dielectric constant along both the [110] and the [$1\bar{1}0$] directions, but did not observe any substantial difference.

[25] We also measured $P_{//}$ in magnetic fields, but $P_{//}$ is by an order of magnitude smaller than $P_{\perp}$, which may be caused by $P_{\perp}$ component due to some misalignment of sample axes.

[26] J. Kanamori, J. Phys. Chem. Solids **10**, 87 (1959).

[27] J. B. Goodenough, *Magnetism and the Chemical Bond*, (Interscience, 1963).

[28] P. W. Anderson, in *Magnetism I*, edited by G. Rado and H. Shuhl (Academic Press, 1963) Chapter 2.

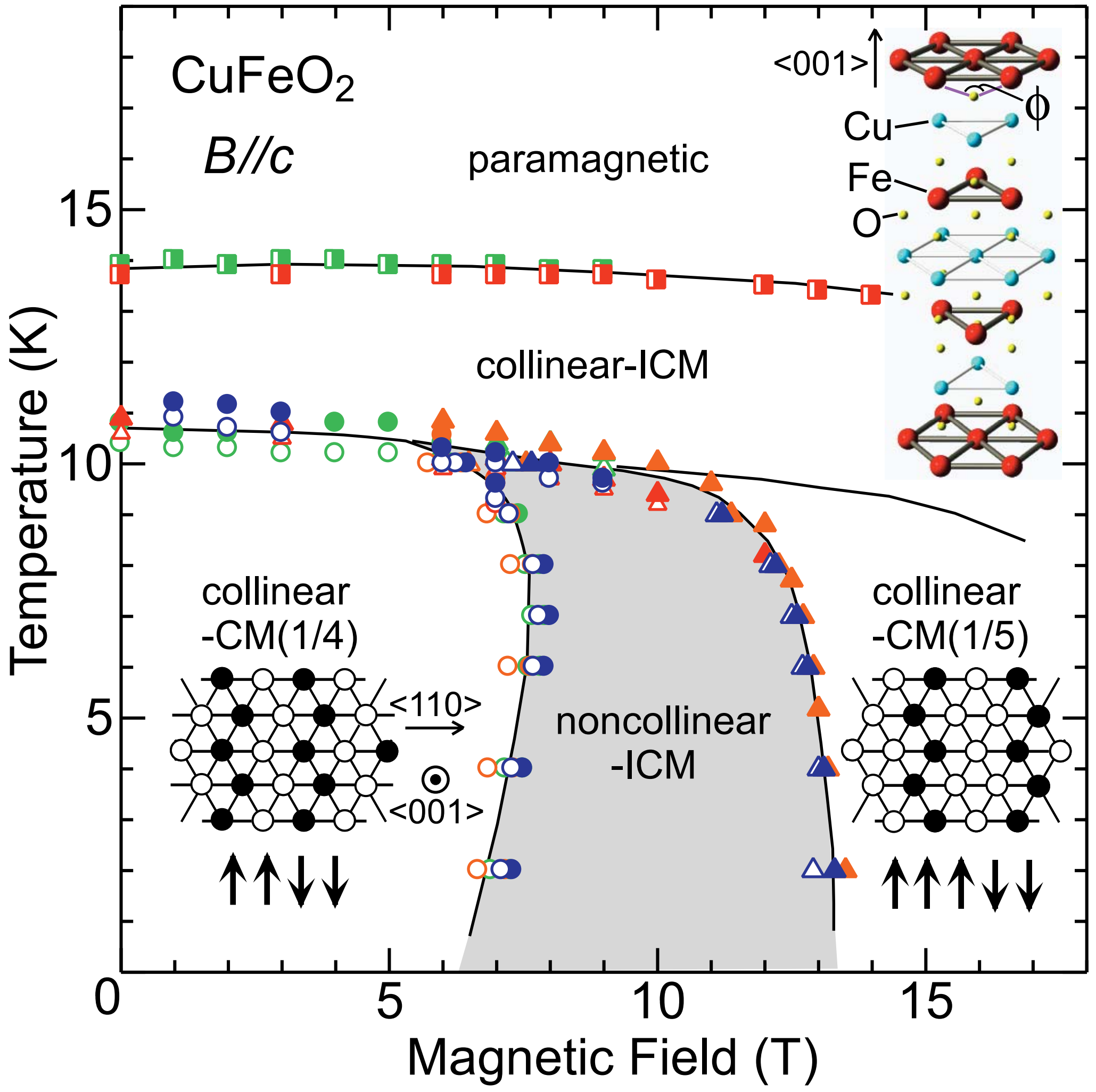

T. Kimura et al. Figure 1

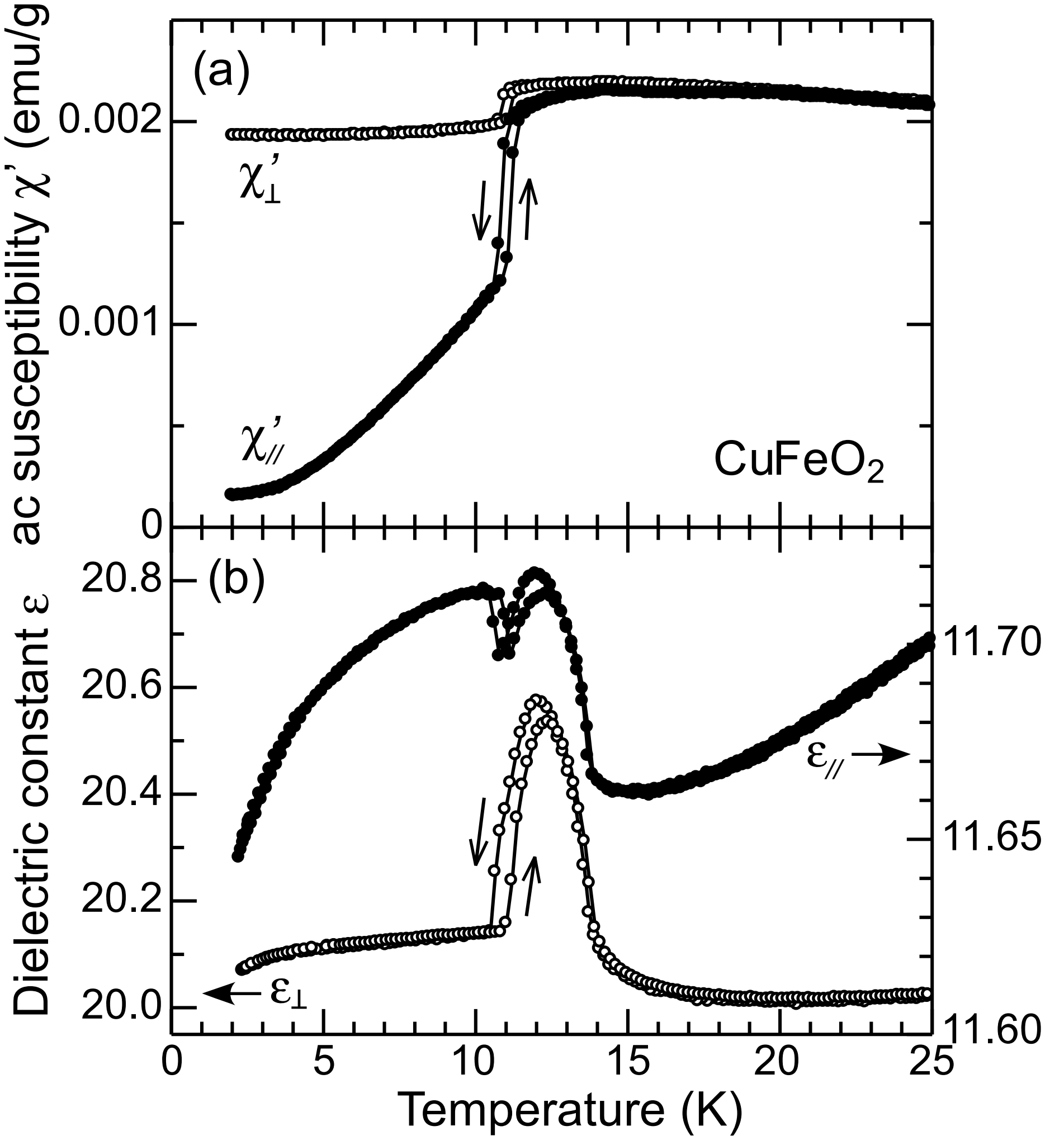

T. Kimura et al. Figure 2

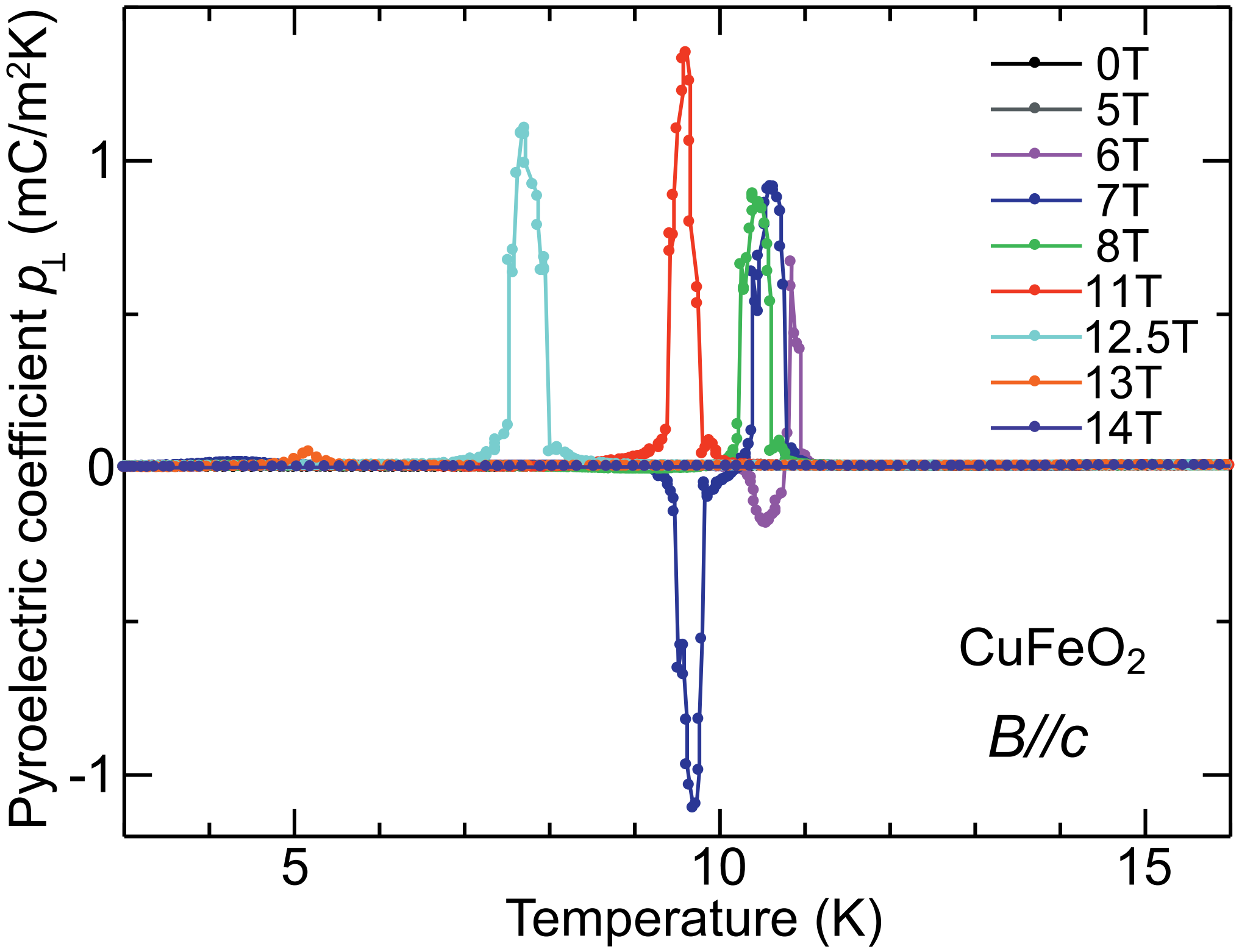

T. Kimura et al. Figure 3

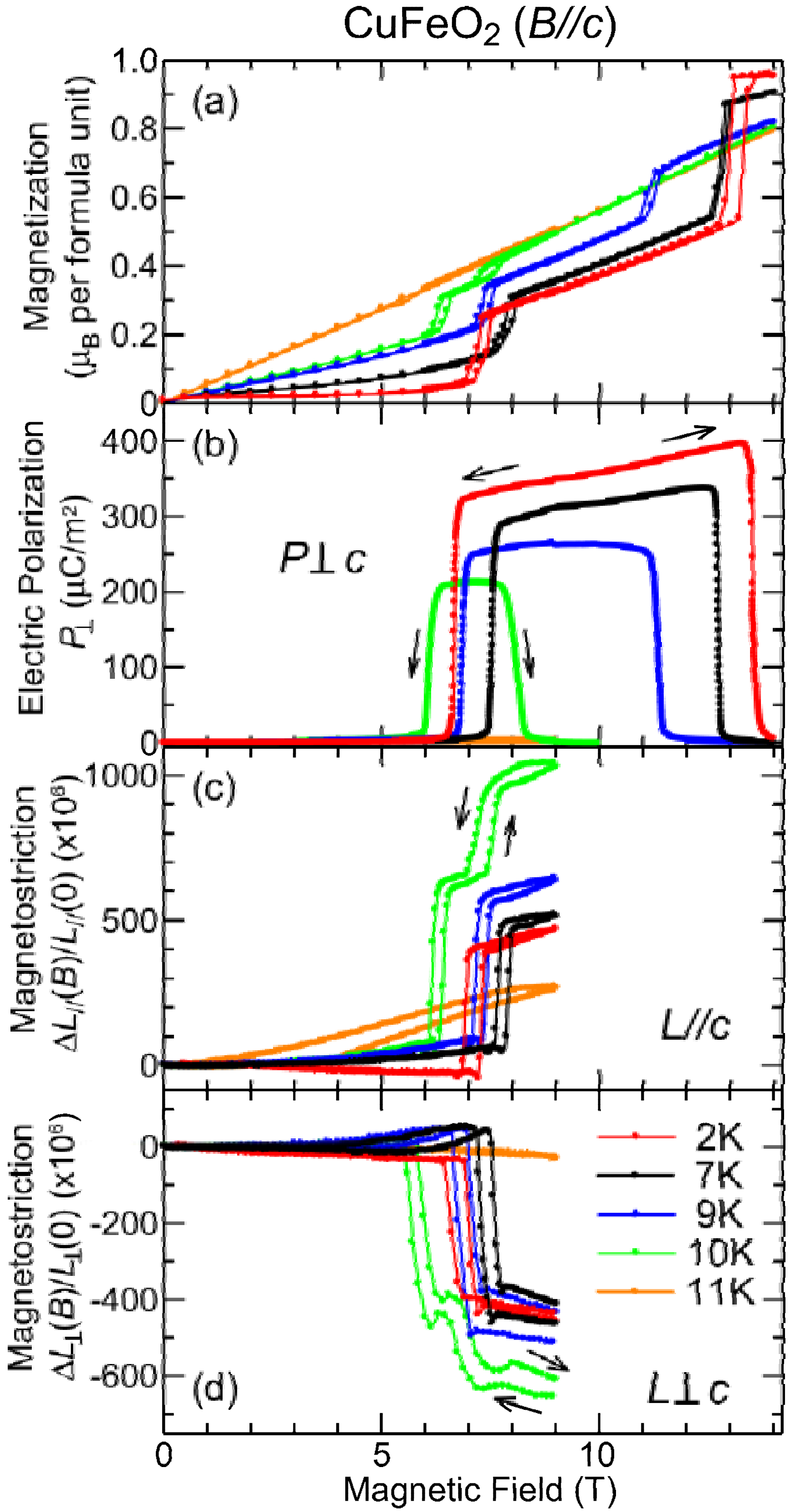

T. Kimura et al. Figure 4